\documentclass[12pt,preprint]{aastex}
\usepackage{amsfonts}
\usepackage{amsmath}
\begin{document}
\title{On the orbital evolution of a giant planet pair embedded in a gaseous disk. II. A Saturn-Jupiter configuration}

\author{Hui Zhang and Ji-Lin Zhou }
\altaffiltext{}{Department of Astronomy \& Key Laboratory of Modern Astronomy
and Astrophysics in Ministry of Education£¬Nanjing University, Nanjing
210093,China ; huizhang@nju.edu.cn}

\begin{abstract}
We carry out a series of high-resolution ($1024\times 1024$) hydrodynamic
simulations to investigate the orbital evolution of a Saturn-Jupiter pair
embedded in a gaseous disk. This work extends the results of our previous work
by exploring a different orbital configuration---Jupiter lies outside Saturn
($q<1$, where $q\equiv M_i/M_o$ is the mass ratio of the inner planet and the
outer one). We focus on the effects of different initial separations ($d$)
between the two planets and the various surface density profiles of the disk,
where $\sigma \propto r^{-\alpha}$. We also compare the results of different
orbital configurations of the planet pair. Our results show that: (1) when the
initial separation is relatively large($d>d_{iLr}$, where $d_{iLr}$ is the
distance between Jupiter and its first inner Lindblad resonance), the two
planets undergo divergent migration. However, the inward migration of Saturn
could be halted when Jupiter compresses the inner disk in which Saturn is
embedded. (2) Convergent migration occurs when the initial separation is
smaller ($d<d_{iLr}$) and the density slope of the disk is nearly flat
($\alpha<1/2$). Saturn is then forced by Jupiter to migrate inward when the two
planets are trapped into mean motion resonances (MMRs), and Saturn may get very
close to the central star. (3) In the case of $q<1$, the eccentricity of Saturn
could be excited to a very high value ($e_{S}\sim 0.4-0.5$) by the MMRs and the
system could maintain stability. These results explain the formation of MMRs in
the exoplanet systems where the outer planet is more massive than the inner
one. It also helps us to understand the origin of the "hot Jupiter/Saturn"
undergoing high eccentric orbit.
\end{abstract}

\section{Introduction}
The existence of more than 40 multiple planet systems have been affirmed so
far. The observational facts show that almost one-fourth of them contain two or
more planets locked in the mean motion resonances (MMRs). This ratio keeps
growing as new detection methods are adopted, e.g. the transit time variation
method which is particularly suitable for detecting low-mass planets locked in
resonance. If we set $q=M_i/M_o$, where $M_i$ is the mass of the inner planet
and $M_o$ is the mass of the outer one, the resonant systems could be simply
divided into two types: (1) $q>1$, e.g. 55 cnc\citep{Fis03}, and (2) $q<1$,
e.g. Gliese 876\citep{Mar01}. The establishments of MMRs when $q>1$ could be
explained by the convergent migration of the planets \citep{Mas01,Mas03}.
According to the theory of disk-planet interaction, planets embedded in a
gaseous disk may undergo various types of migration depending on their masses.
Low(around several Earth masses) or moderate mass(around Saturn mass) planets
usually undergoes fast type I or III migration, while massive planets
(comparable to Jupiter) opens a gap on the gaseous disk and undergoes slow type
II migration. In the case of 55 cnc, for example, the inner planet has a
minimum mass of $0.87M_J$ and the outer one has a minimum mass of $0.17M_J$,
where $M_J$ is the Jupiter mass. For this mass configuration, the outer planet
probably migrates inward faster than the inner one and catches the inner one in
the $3:1$ MMR. These kinds of processes have been explored numerically by
several classic works \citep{Sne01,Nel02,Pap05,Kle04,Kle05}, as well as our
previous research(Zhang \& Zhou 2010, hereafter Paper I). Although the process
is complicated, the convergent migration and the establishment of MMRs are
robust outcomes.

If the above-mentioned mechanism also dominates the orbital evolution when
$q<1$, then divergent migration should occur naturally and the establishments
of MMRs would become ineffective. However, observations show that the MMRs
exist in exoplanet systems of this orbital configuration as well.
\textbf{Table} \ref{table 1} shows a list of exoplanet systems that contain two
planets probably locked in MMRs. In fact, one may find that more than half of
the resonant systems have $q<1$. How do these MMRs form? There should be a
mechanism that suppresses or even halts the fast inward migration of the
low-mass inner planet, so that the massive outer one could catch it before it
is swallowed by the star. If such a mechanism exists, it may also help us to
better understand the formation of a "hot Jupiter/Saturn" which survives the
inward migration.

For terrestrial planets (tens of Earth masses or below), their inward migration
could be halted by the density jump on the disk \citep{Mas06,Mor08}. Otherwise,
they could be captured by the high-order MMRs of an inward migrating
giants\citep{Zho05,Fog07,Ray06}. These processes explain the formation of
terrestrial planets in "hot Jupiter/Saturn" systems well, but could not account
for the low-order MMRs in the systems whose inner planet is as massive as
Saturn or even Jupiter, e.g. Gliese 876, HD 160691 and HD 128311.

So far, few works have considered the orbital configuration in which the two
\emph{giant} planets have a mass ratio $q<1$. Kley et al. (2005) studied the
orbital evolution of Gliese 876, which has $q\approx0.31$. They simulated
convergent migration for the two planets and successfully represented the their
observed orbital configuration, especially for the right range of
eccentricities. To obtain convergent migration, they assumed that there is a
preformed cavity at the center of the disk. The two planets are set inside the
cavity so that the outer planet keeps losing its angular momentum to the disk
and migrates inward, while the inner planet barely migrates until it is trapped
by MMRs with outer one. This assumption in fact implies that gas accretion onto
the center star should be so vigorous that the inner part of the disk is
depleted much earlier before the formation and migration of the giant planets.
Other processes like the magnetic field may also account for the inner cavity
at the center of the disk. However, the relatively small radii of the cavity
limit this assumption to specific cases. A more general and self-consistent
mechanism to bring about convergent migration is required in a whole and
regular disk. To construct a full-region disk instead of a ring, we adopt the
Cartesian computational domain. Details of these domain settings can be found
in paper I.

As shown in \textbf{Table} \ref{table 1}, we note that most of the planet pairs
trapped in MMRs have $q\approx 0.3-0.6$. A familiar and typical planet pair in
this range is a Jupiter-Saturn pair, which is most suitable for studying the
dynamics of this mass configuration. Therefore, following Paper I, we continue
to investigate the orbital evolution of a Saturn(inner)-Jupiter(outer) pair
embedded in a gaseous protostellar disk. We focus on the effects of various
surface density slopes($\alpha=-d \ln (\sigma)/ d \ln (r)$) and various initial
separations $d$ between the two planets. We will show the following. (1)
Although under divergent migration, when Jupiter migrates inward, it could halt
the inward migration of Saturn by compressing the inner disk in which Saturn is
embedded. (2) When the initial separation is smaller than the distance from
Jupiter to its farthest inner Lindblad resonance $d_{iLr}$ and the density
slope of the disk is nearly flat, the two planets may undergo convergent
migration and then be trapped into MMRs. (3) The eccentricity excitation of
MMRs overwhelm the damping of the gas disk. Saturn may get very close to the
central star($a_s<1$AU) preserving a very high eccentricity. (4) We also
compare the results between different orbital configurations---$q>1$ and $q<1$.
These results help reveal the orbital architecture formation of some resonant
exoplanet systems, e.g. HD 160691 and Gliese 876, as well as the formation of
the 'hot Jupiter/Saturn' undergoing high eccentric orbit.

This paper is organized as follows: the model including the numerical methods
and the initial settings are introduced in Section 2; our results and the
analysis are presented in Section 3; some discussions are made in Section 4,
and the conclusions are summarized in Section 5.

\section{Numerical setup}
\subsection{Physical model}
We simulate the full dynamical interaction of a system including a solar-type
star, a Saturn mass giant(inner planet), a Jupiter mass giant(outer planet) and
a two-dimensional (2D) gas disk. The star is fixed at the origin of the system
with both the planets and the disk surrounding it; thus, the whole system is
accelerated by the gravity of the planets and disk. For efficiency, we ignore
the self-gravitating effect of the gas. Therefore, the gravity exerted on the
gas only comes from the central star, the two giant planets and the
acceleration of the origin.

For numerical convenience, the gravitational constant $G$ is set to $1$. The
solar mass ($M_{\odot}$) and the initial semimajor axis of Saturn ($R_{s0}=5.2$
AU) are set in units of mass and length, respectively. We locate Saturn
initially inside the orbit of Jupiter; thus, the mass ratio of the two planets
is $q\approx0.33$. Our aim is to investigate the formation of MMRs and the
consequential evolution of the system with this orbital configuration.
Considering that Saturn and Jupiter would mostly undergo type II migration, we
adopt a relatively large viscosity of $\nu=5\times10^{-5}$ to accelerate the
evolution, and run simulations up to $2500-5000$ initial orbital periods of the
inner planet.

We adopt a polytropic equation of state and set the disk aspect ratio to be
$H/r=0.04$. The evolution of the gas under the gravity of the star and planets
is solved by the 2D Godunov coded \emph{Antares}, which is based on the exact
Riemann solution for isothermal or polytropic gas. While the dynamics of the
two planets under the potential of the star and gas disk are integrated by an
eighth-order Runge-Kutta integrator, the global time step is the minimum of the
hydrodynamical part and the orbit integration part. Details of the numerical
method as well as the computational configuration have been well described in
our previous works(Paper I). A comparison with other well-studied codes has
also been presented in paper I.

\subsection{Initial condition}
One of the issues that we expect to figure out is the effect of the surface
density profile on the disk. In this paper, we try several typical density
profiles, which are only the function of disk radii $\sigma=\sigma_0
r^{\alpha}$. As shown in \textbf{Table} \ref{table 2}, the initial density
distribution varies from flat to very steep: $\sigma_0$, $\sigma_0 r^{-1/2}$,
$\sigma_0 r^{-1}$ and $\sigma_0 r^{-3/2}$. $\sigma_0$ is set to be 0.0006 in
our units, which corresponds to a height-integrated surface density $\sim 200
g/cm^2$. The density slope on gas disk results in a pressure gradient. To
ensure that there is no radial flow at the beginning, we set the radial
velocity $u_{r0}$ to be $0$ and adjust the initial angular velocity of gas
$u_{\theta0}=r \Omega_g$ to balance the pressure and central gravity, where
$\Omega$ depends on $\alpha$.

Another issue is the role of the initial separation between Saturn and Jupiter.
As shown in \textbf{Table} \ref{table 2}, we choose three separations: $d\equiv
a_{J0}-a_{S0} = 1, 0.5, 0.25$. For numerical convenience, we set the initial
semi-major axis of the inner planet (Saturn) as the length unit (the initial
location of Saturn is always $a_{S0}=1$). Then we adjust the initial locations
of the outer one (Jupiter) to obtain different separations. When $d=1$, Jupiter
initially locates at $a_{J0}=2$. At such a large distance, the mutual
interaction due to gravity is negligible, so the two planets could migrate
independent of each other at the very beginning. When $d=0.5$, Jupiter is
initially located at $a_{J0}=1.5$. The position of the $P_J:P_S=2:1$ MMR, where
$P_J$ and $P_S$ are the orbital periods of Jupiter and Saturn, respectively is
now at $r\approx0.94$, which is a little bit inside the initial location of
Saturn. By doing so, Saturn passes through the $2:1$ resonance of Jupiter soon
after their release if divergent migration occurs. Furthermore, the first inner
Lindblad resonance (at $\Omega=\Omega_J+\kappa/m$, where $\kappa$ is the
epicycle frequency and $m=-2$) of Jupiter is located around $\sim0.88-0.94$
depending on the density slope $\alpha$. Now the separation between the two
planets is smaller than the distance between Jupiter and its first(also the
furthest) inner Lindblad resonance, $d\lesssim d_{iLr}$. Thus, the inner
Lindblad torque of Jupiter should be reduced by the existence of Saturn at the
beginning of evolution. Finally, when $d=0.25$, Jupiter initially locates at
$a_{J0}=1.25$. This location ensures that Saturn passes through the
$P_J:P_S=3:2$ MMR of Jupiter, the location of which is at $r=0.95$. More
importantly, this small separation $d \leq 3 r_{mH}$ allows mutual scattering
due to gravity between the two planets at the beginning of evolution, where
$r_{mH}=0.085$ is the mutual Hill radius of Saturn and Jupiter at this
configuration:
\begin{equation}
r_{mH} \equiv (\frac{M_J+M_S}{3M_\odot})^{\frac{1}{3}}(\frac{a_J+a_S}{2}).
\end{equation}

The initial settings of the disk do not take into account of the gravitational
perturbation of the planets. Instead, we adopt the "quiet-start" prescription
to setup a dynamical equilibrium to ensure that the streamlines of the gas are
always closed when the planet is growing. Basically, we set the initial masses
of the planets to be negligible, then we fix their orbits and increase their
masses to Saturn and Jupiter masses adiabatically. At the end of their growth
we release them at the same moment and start the evolution. Details of this
prescription have also been described in Zhang et al. (2008).

\section{Results}
\subsection{Divergent migration and suppression of inward migration}
Our main results are summarized in \textbf{Table} \ref{table 2}. There are two
main variables: the density slope $\alpha$ and the initial separation $d$
between the two planets. To avoid confusion, we present the results mainly
according to the sequence of the initial separations $d$. And for each $d$, we
present the effects of the different density slopes first and then explain
these effects together.

\begin{figure}
\epsscale{1} \plotone{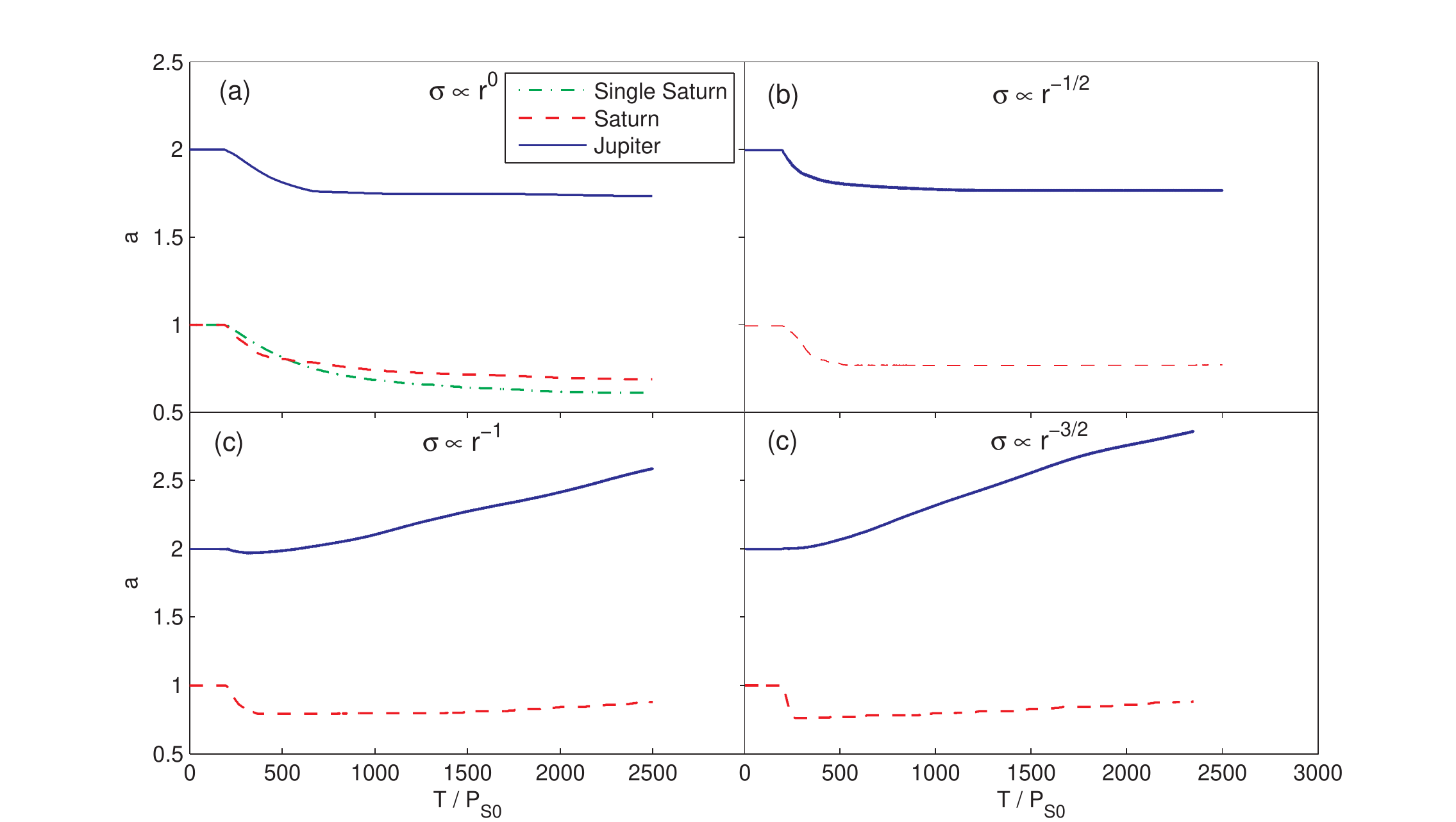} \caption{semi-major axis
evolutions of Saturn and Jupiter embedded in a gas disk whose surface density
slope varies from flat($\sigma\propto r^0$) to very steep($\sigma\propto
r^{-3/2}$). The dash-dot curve in Panel (a) corresponds to the migration of
single Saturn, which is faster than that in planet pair case. This indicates
that the inward migration of Saturn is suppressed by the existence of Jupiter.
Further more the inward migration may be halt(Panel b) or even reversed(Panel
c, d) as the surface density becomes steeper. \label{m1}}
\end{figure}

First, we start with a relatively large initial separation ($d=1$). As
introduced in the previous section, the mutual interaction due to gravity
between Saturn and Jupiter is now negligible ($d>5r_{mH}$). Our results show
that, at large separation, Saturn and Jupiter will generally undergo divergent
migration or achieve equilibrium state, depending on the surface density
profile:

(1) When the disk is very steep ($\alpha>1/2$), Jupiter digs a clear gap and
migrates outward after the release. In the meantime, Saturn migrates inward
very fast until it clears its coorbital region(type III migration). Then Saturn
starts to follow the viscous evolution of the disk, and its fast inward
migration is halted or even reversed. This result is consistent with our
previous result that the massive planet follows the movement of the global disk
and moves outward when $\alpha>1/2$. \textbf{Figure} \ref{m1} shows the orbital
evolutions of the two giant planets embedded in different surface density
slopes when $d=1$.

(2) When the disk is nearly flat ($\alpha\leq1/2$), both planets are under
inward migration. Since Saturn is still surrounded by gas at the moment of
release, it migrates inward much faster than Jupiter during the first stage of
evolution. As Saturn keeps clearing its vicinity, inward migration is then
reduced gently. In fact, we observed an equilibrium state where Jupiter and
Saturn both stop migrating inward and maintain their separation when
$\alpha=1/2$, see Panel (b) in \textbf{Figure} \ref{m1}. This is also
consistent with the results of Paper I, which showed that the direction of
viscous movement changes sign around $\alpha=1/2$:
\begin{equation}
\dot{r}=\frac{1}{2\pi r \sigma}\frac{\partial\Gamma_\nu/\partial
r}{d(r^2\Omega)/dr}= -3\nu(\frac{1}{2}-\alpha)r^{-1}. \label{rvis}
\end{equation}
where $\Gamma_\nu$ is the viscous torque,
\begin{equation}
\Gamma_\nu=2\pi r^2 \nu \sigma  \frac{r d\Omega}{dr} \sim r^{1/2-\alpha},
\end{equation}
which stays constant over different radii $r$ when $\alpha=1/2$, by assuming a
constant viscosity $\nu$ across the disk.

\begin{figure}
\epsscale{1} \plotone{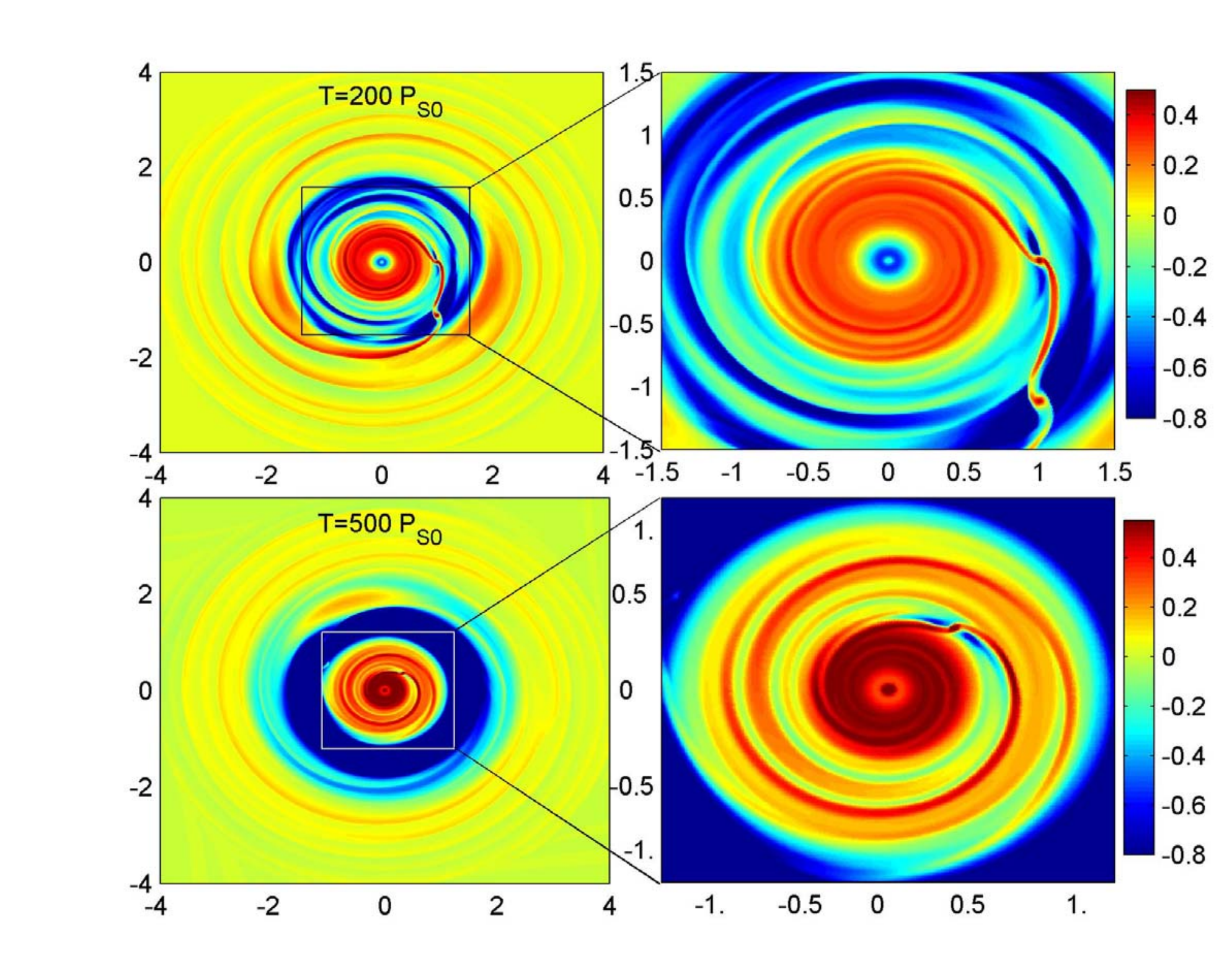} \caption{density map of the
gas disk at different evolution stages, where the initial density slope
$\alpha=0$ and the initial separation $d=1$. Note that, after about $500P_{S0}$
evolution from release, the inner disk is compressed and shepherded by Jupiter
while Saturn is digging a gap on it. This ensures Saturn moves with the inner
disk. \label{d01}}
\end{figure}

\begin{figure}
\epsscale{0.9} \plotone{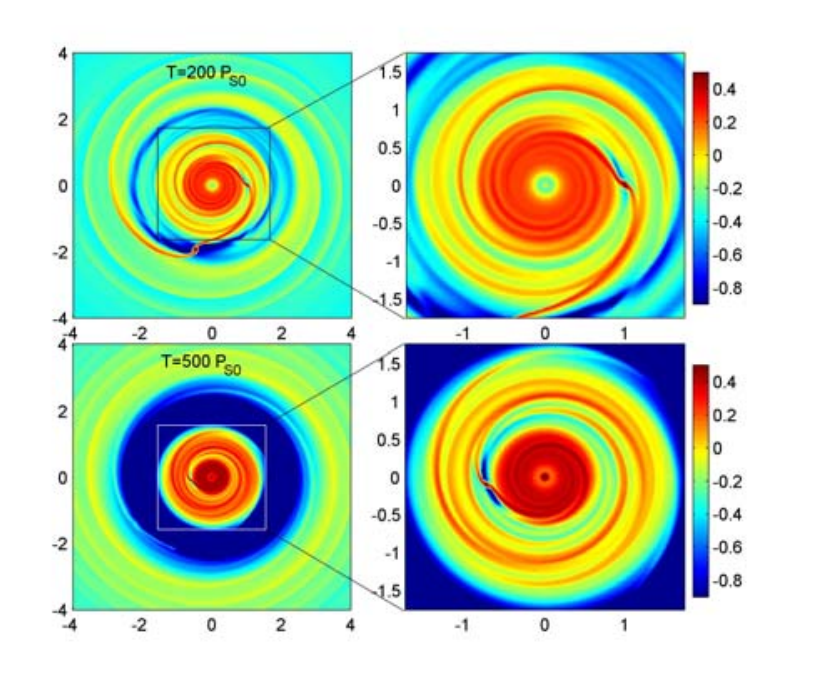} \caption{density map of
the gas disk at different evolution stages, where the initial density profile
is steeper than \textbf{Figure} 2, $\alpha=1/2$ and the initial separation is
still $d=1$. This steeper density slope prevents the inward migration of
Jupiter, as well as that of Saturn. \label{d051}}
\end{figure}

According to \textbf{Figure} \ref{m1}, one may find that both the inward and
the outward type II migration of Saturn are suppressed. After several hundred
orbits evolution, the whole disk has been well separated into two parts---an
inner disk and an outer disk---by the gap opened by Jupiter. Saturn has also
dug a gap in the inner disk. Although it is much weaker than that of Jupiter,
this gap ensures that Saturn will undergo type II migration. Because of the
large initial separation and divergent migration at the beginning stage, the
gaps of the two planets will not overlap soon. Thus, the inner disk and Saturn
could be treated as a sub-system which is shepherded by the tidal torque of
Jupiter, see \textbf{Figures} \ref{d01} and \ref{d051}. Then the surface
density profile $\alpha$ makes some differences.

\begin{figure}
\epsscale{1} \plotone{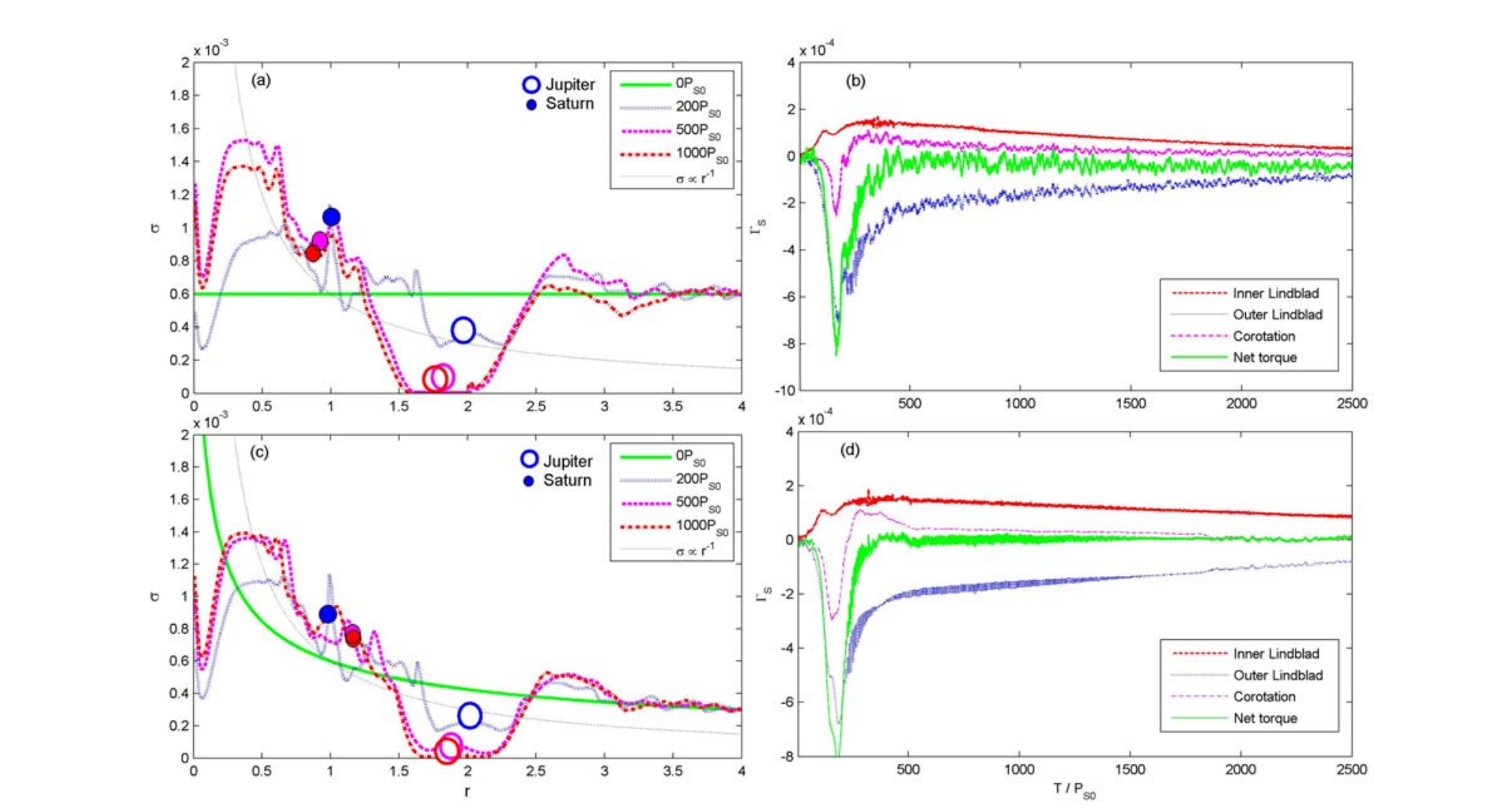} \caption{cross sections of the
gas disk at different stages and the associated torque evolutions. The initial
separations are $d=1$. Panel (a): density cross sections at $T=0$, $200P_{S0}$,
$500P_{S0}$ and $1000P_{S0}$, where $\alpha=0$. The light doted line shows the
distribution of $\sigma\propto r^{-1}$ as a reference. Panel (b): the evolution
of torques exerted on Saturn, when $\alpha=0$. Panel (c): density cross
sections at $T=0$, $200P_{S0}$, $500P_{S0}$ and $1000P_{S0}$, where
$\alpha=1/2$. The light doted line shows the distribution of $\sigma\propto
r^{-1}$. Panel (d): the associated evolution of torques exerted on Saturn when
$\alpha=1/2$. Note that the surface density increases at the inner disk and its
local surface density becomes steeper. The outer Lindblad torque is initially
much larger than the inner one, then it decreases very quickly while the inner
one increases a bit after the release. And the corotation torque changes from
negative to positive as some gas is forced to flow across Saturn's orbit from
outside to inside. Thus the net torque is greatly negative at beginning and
soon increases to near or even equal to zero. \label{s01051}}
\end{figure}

If the disk was nearly flat ($\alpha\leq1/2$), Jupiter migrates inward gently
and pushes the gas of the inner disk toward the central star. Thus the local
surface density distribution of the inner disk is changed. As shown in
\textbf{Figure} \ref{s01051}, the surface density profile of the inner disk is
changed from flat ($\alpha\leq1/2$) to relatively steep ($\alpha\geq1$). When
the local disk mass exceeds the planet mass, $\pi a_p \sigma\gtrsim M_p$, the
migration of the planet is then disk dominated. To exceed the mass of Saturn,
the disk requires a minimum local density $\sigma_{min}\geq 1.5\times10^{-4}$
inside the orbit of Saturn when Saturn is located at $a_S=0.8$, which is much
smaller than the present density of the inner disk ($\sigma>6\times10^{-4}$).
This ensures that the type II migration of Saturn is dominated by the disk
evolution, which means Saturn follows the movement of the inner disk. On the
one hand, the gas which flows across the orbit of Saturn from outside to inside
exerts an additional positive corotation torque on Saturn, see Panel (b) and
(d) in \textbf{Figure} \ref{s01051}. On the other hand, as the inner part
becomes denser, the inner disk tends to spread outward. However the outward
diffusion of the inner disk is suppressed by Jupiter(as well as the disk
outside Jupiter's orbit), thus this local density profile is maintained(Panel
(a) and (c) of \textbf{Figure} \ref{s01051}). As a result, the inward migration
of Saturn is slowed down or even halted.

If the disk is relatively steep ($\alpha>1/2$), Jupiter tends to migrate
outward. The inner disk could now expand outward, and Saturn follows the
movement of the gas. However, the Hill radius of Jupiter increases as it moves
outward and the expansion of the inner disk is limited by the increasing width
of the gap. Thus the outward migration of Saturn is also slower than that of
Jupiter, see Panel (c) and (d) in \textbf{Figure} \ref{m1}.

The above results indicate that Saturn could migrate slower than Jupiter under
some conditions, when they are both undergoing type II migration. If the
surface density of disk is steep, both Saturn and Jupiter will migrate outward
and thus, results in divergent migration, while in a nearly flat disk, the
inward migration of Saturn is suppressed. Due to the large initial separation,
the two planets would achieve an equilibrium state(Panel (a) and (b) in
\textbf{Figure} \ref{m1}). How could they then approach to each other further
and get into MMRs? One possible way is reducing their initial separation to
enlarge the interactions between them at the very beginning of the evolution.

\subsection{convergent migration and MMR captures}
\begin{figure}
\epsscale{1} \plotone{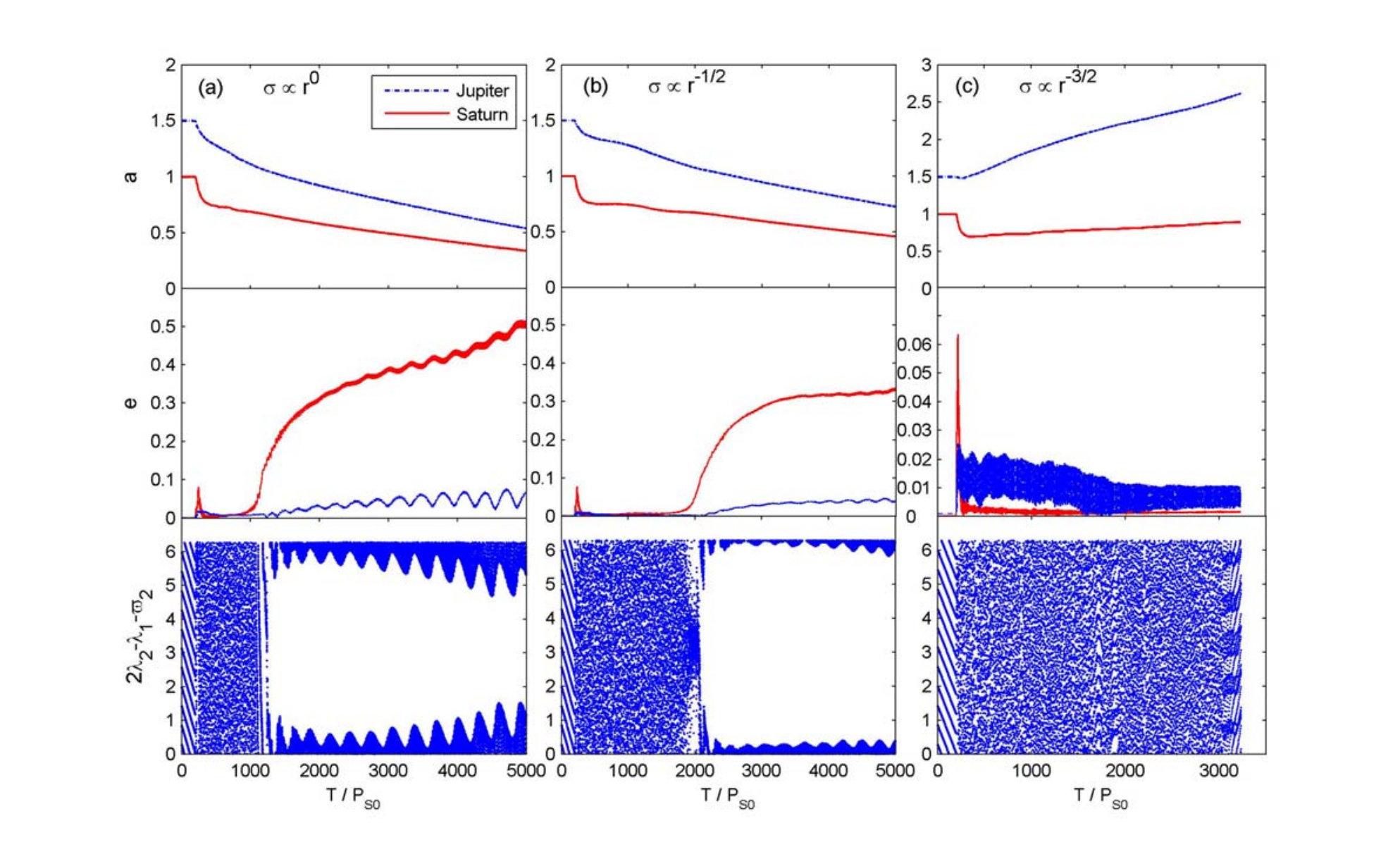} \caption{orbital
evolutions of Saturn and Jupiter embedded in a gas disk whose surface density
slope $\alpha$ is $0$(Panel a), $1/2$(Panel b) and $3/2$(Panel c). The initial
separation between the two planets is reduced to $d=0.5$. Convergent migration
happens when the disk is nearly flat $\alpha\leq1/2$. \label{m05}}
\end{figure}

Next, we try a moderate separation $d=0.5$ by setting Jupiter at $a_{J0}=1.5$
and Saturn at $a_{S0}=1$. At this distance, the initial mutual interaction due
to the gravity of the two planets is still not important ($d>4r_{mH}$). However
the indirect interaction becomes significant since Saturn initially locates
near the position of the $2:1$ MMR with Jupiter, and furthermore, it is also
close to the location of the farthest inner Lindblad resonance ($m=-2$) of
Jupiter. Our results show that the surface density profile also plays a great
role in the final results:

(1) When $\alpha>1/2$, we get divergent migrations which are similar to the
cases with $d=1$. The divergent rate is higher when a larger $\alpha$ is
adopted. An important difference is that Jupiter migrates inward for a while
right after the moment of release and turns back outward while Saturn had moved
far away inward, see Panel (c) in \textbf{Figure} \ref{m05}. This phenomenon
indicates that, at small separation, the total inner Lindblad torque exerted on
Jupiter is weakened by Saturn(at the beginning of evolution).

(2) When $\alpha\leq1/2$, we find that Jupiter migrates inward much faster than
it does in the $d=1$ cases, and the inward migration of Saturn is substantially
slowed down. At this condition, we observe a gently convergent migration
between Saturn and Jupiter. This slow convergent migration then results in the
$2:1$ MMR of the two planets and Saturn is forced to migrate further inward by
Jupiter. Their eccentricities are greatly excited and maintained by the
resonance, especially for that of Saturn ($e_{S}\sim 0.4-0.5$), see Panel (a)
and (b) in \textbf{Figure} \ref{m05}.

\begin{figure}
\epsscale{1} \plotone{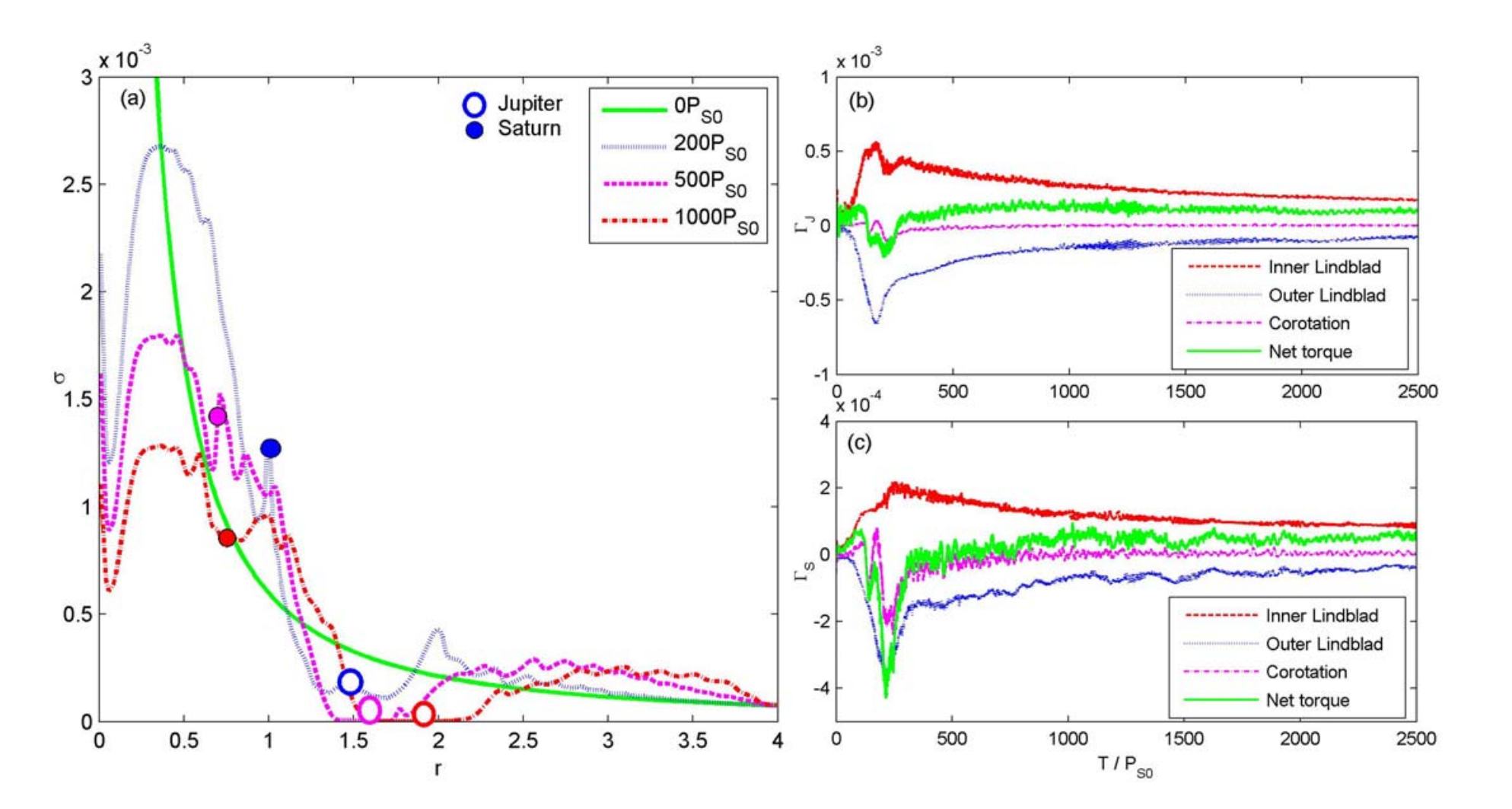} \caption{Panel (a): the surface
density evolution when $\alpha=3/2$. Panel (b): the evolution of torques
exerted on Jupiter. Note that the jump around $T=250P_{S0}$ indicates the
recovery of the inner Lindblad torque as Saturn migrates further away. Although
the outer Lindblad torque is initially larger, the inner Lindblad torque
decreases slower than the outer one does and makes the net torque positive soon
after the release. Panel (c): the evolution of torques exerted on Saturn. Note
that the corotation torque decreases to zero as Saturn digs a gap on the inner
disk and the outer Lindblad torque decreases much faster than the inner one.
\label{s1505}}
\end{figure}

\begin{figure}
\epsscale{1} \plotone{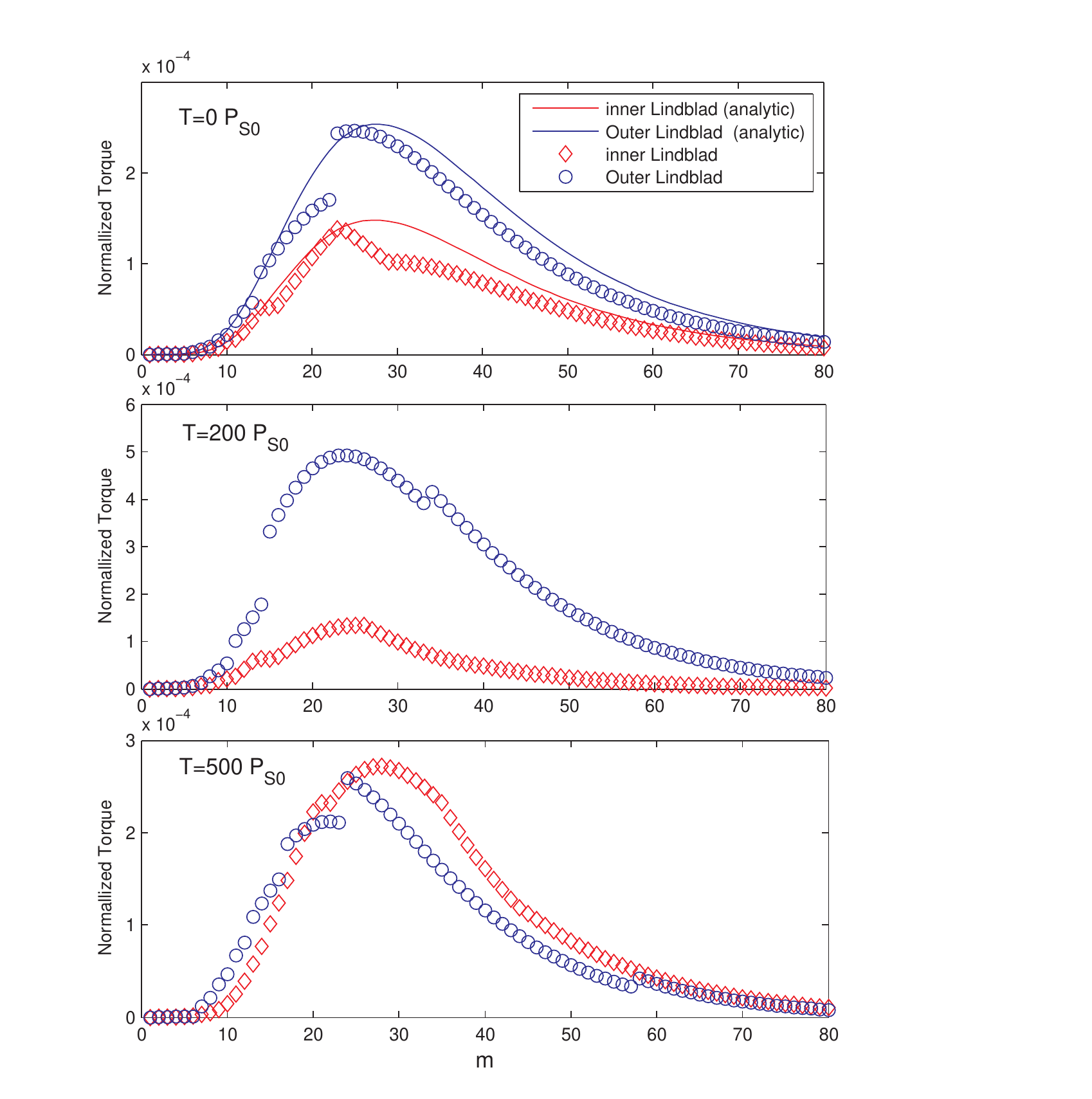} \caption{semi-analytic
results of the inner and outer Lindblad torques exerted on Saturn, for each
integer $|m|=2\sim80$. The density slope $\alpha=3/2$ and the initial
separation $d=0.5$. The maximum of inner torque increases and moves slightly
toward high value of $m$, this indicates the compression(surface density
increases) and outward expands of the inner disk(the inner edge of gap which
opened by Saturn moves toward Saturn). As the outer Lindblad torques decreases
faster, the net torque becomes positive at around $T=500P_{S0}$. \label{l1505}}
\end{figure}

These results can be understood as follows. At the beginning of the evolution,
Saturn sweeps out the gas at the positions of the inner Lindblad resonances of
Jupiter. Thus, the total inner Lindblad torque exerted on Jupiter is weakened.
The outer Lindblad torque then dominates the migration of Jupiter for a while.
When $\alpha>1/2$, their orbits become divergent because Saturn migrates inward
faster than Jupiter at the first stage of the evolution. When their separation
becomes large enough, Saturn's effect vanishes. Thus, the inner Lindblad torque
overwhelms the outer one again and Jupiter start to migrate outward. Panel (b)
in \textbf{Figure} \ref{s1505} shows the torques exerted on Jupiter versus time
when $\alpha=3/2$. It is clear that the inner torque decreases for a while
around the moment of release. Then the inner torque recovers and overwhelms the
outer one, while Saturn is quickly dragged inward by the corotation torque at
the beginning (Panel (c) of \textbf{Figure} \ref{s1505}). Then, as its
coorbital region is cleared, Saturn follows the expansion of the inner disk and
moves outward. \textbf{Figure} \ref{l1505} shows the semi-analysis of each
$m$th Lindblad torque (inner and outer) exerted on Saturn(obtained by the same
method in paper I). One may find that the total outer Lindblad torque decreases
a lot after the release while the inner one increases a little bit, and thus
the net torque becomes positive.

\begin{figure}
\epsscale{0.9} \plotone{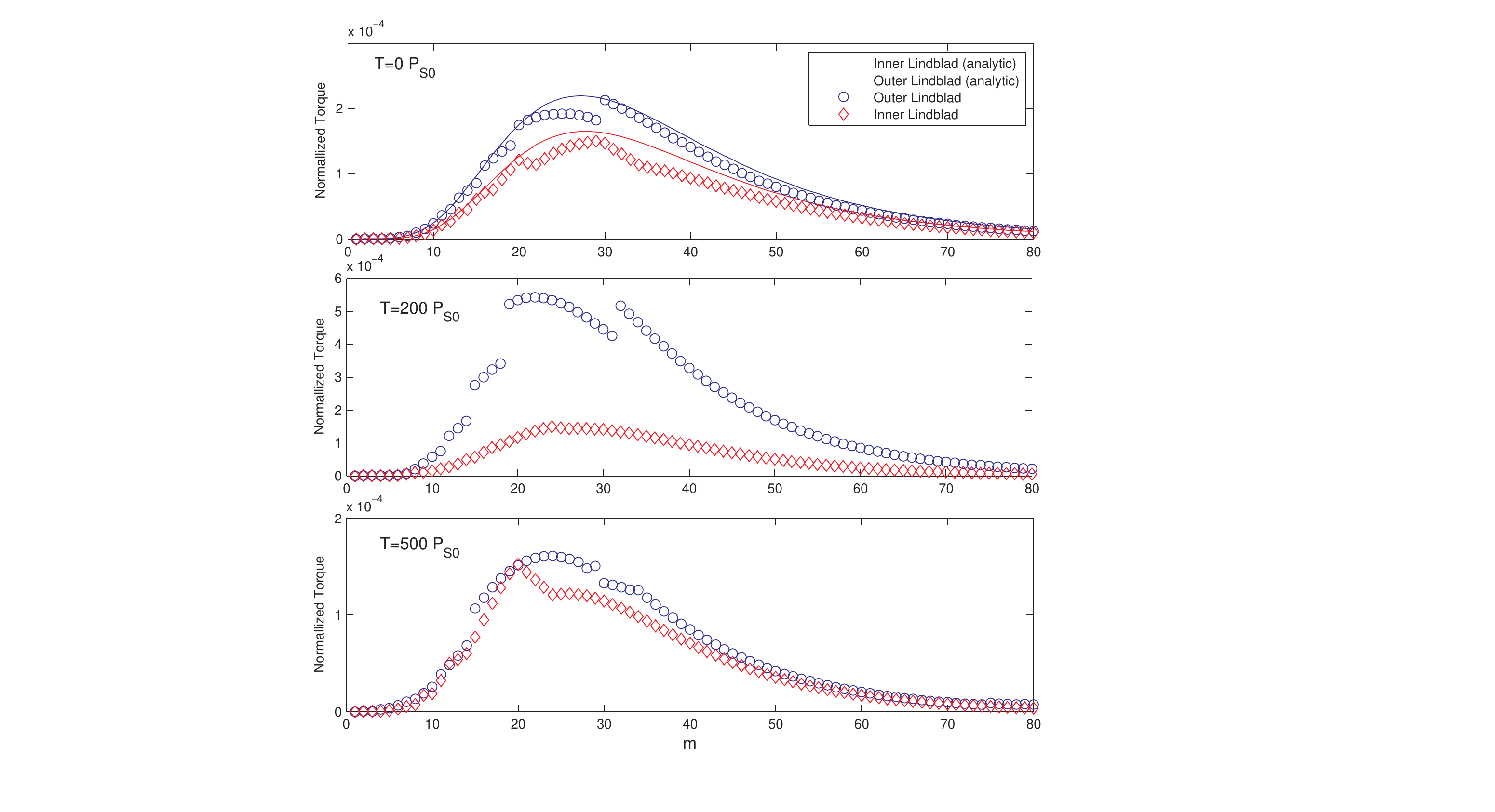} \caption{semi-analytic
results of the inner and outer Lindblad torques exerted on Saturn, for each
integer $|m|=2\sim80$. The density slope $\alpha=0$ and the initial separation
$d=0.5$. The outer Lindblad torques decrease while the inner ones maintain.
Both the maximum of inner and outer torques move toward small $m$, which
indicates a gap around Saturn is forming(surface density decreases around
Saturn). \label{l005}}
\end{figure}

\begin{figure}
\epsscale{0.9} \plotone{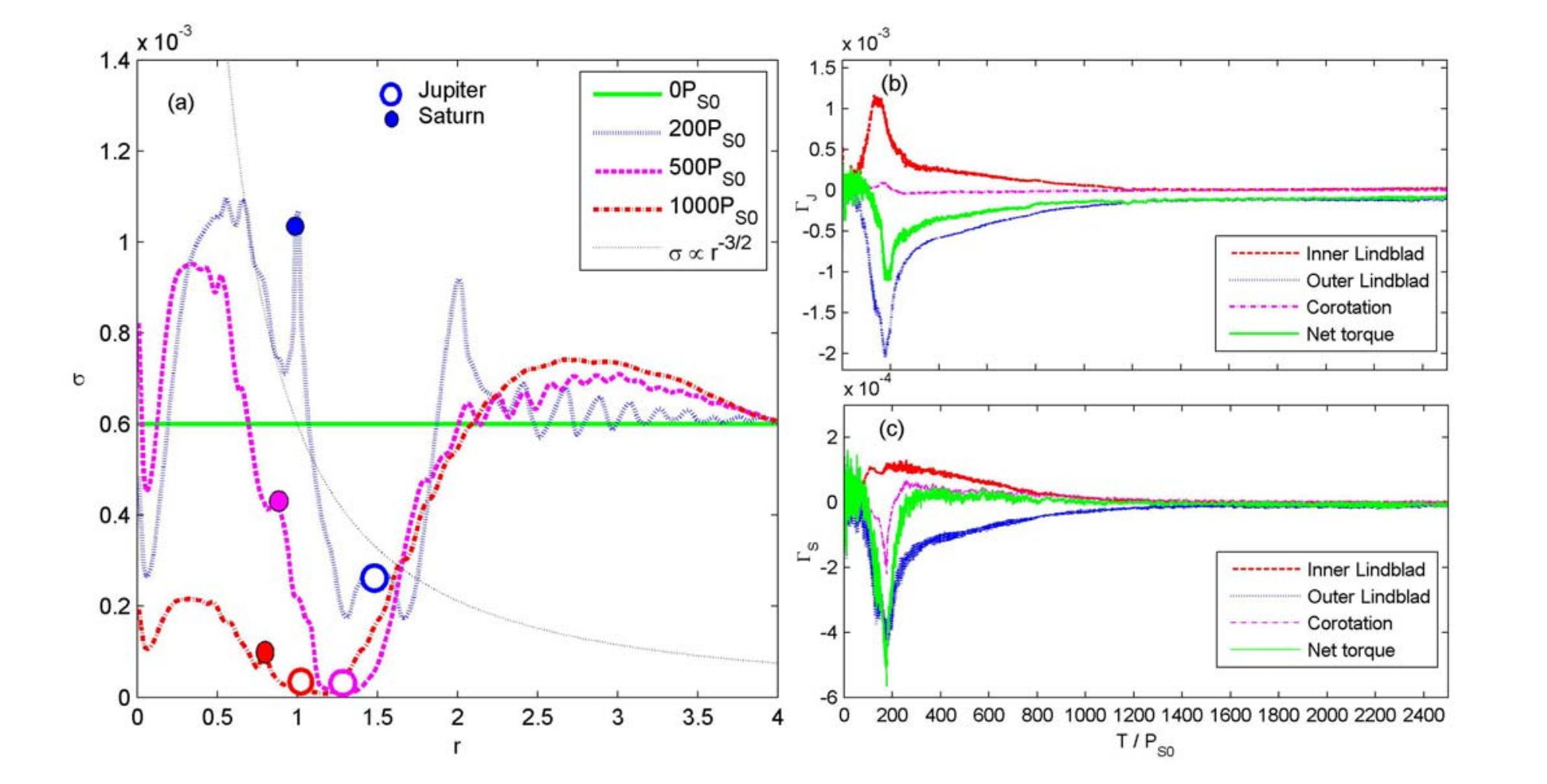} \caption{Panel (a): surface
density evolution when $\alpha=0$ and $d=0.5$. The inner disk is swept out by
the inward migrating planet pair and a cavity forms at $T\geq1000P_{S0}$. Panel
(b): the evolution of torques exerted on Jupiter. Panel (c): the evolution of
torques exerted on Saturn. Note that the outer Lindblad torque decreases much
faster than the inner one after the release moment. \label{s005}}
\end{figure}

\begin{figure}
\epsscale{0.9} \plotone{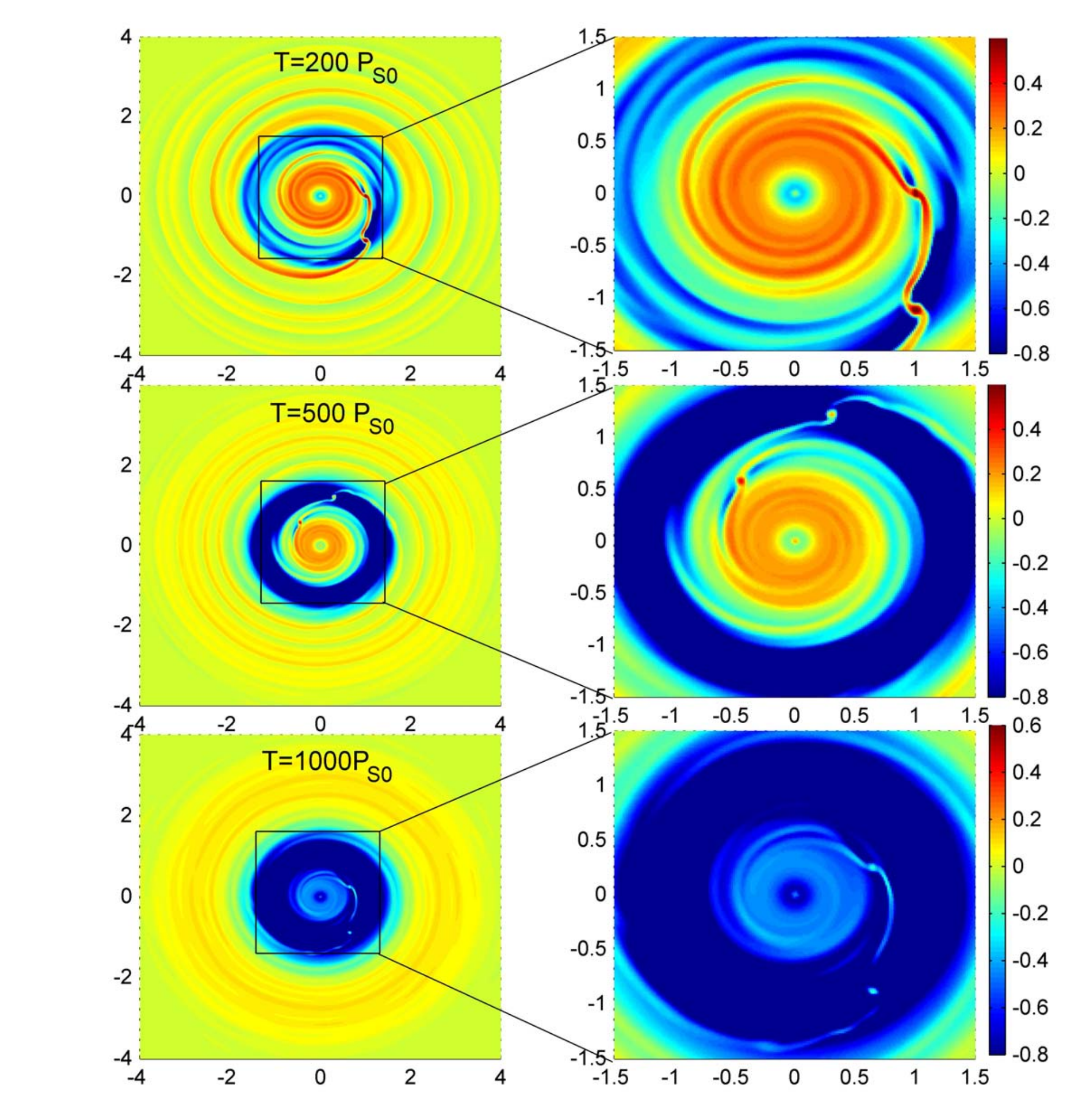} \caption{density map at
different stages of evolution. The right Panels zoom in the inner disk in which
Saturn is embedded. \label{d005}}
\end{figure}

The situation is different when $\alpha\leq1/2$. Although the decrease rate of
the inner Lindblad torque is slowed down when the separation between Saturn and
Jupiter is increasing, the outer Lindblad torque still dominates the migration
of Jupiter and pushes it inward slowly. In the meantime, the inward migration
rate of Saturn decreases even more. This is mainly due to three reasons. First,
as we have mentioned before, the inward migrating Jupiter compresses the inner
disk and the steepened local surface density profile makes the net torque
exerted on Saturn vanish there. \textbf{Figure} \ref{l005} shows each $m$th
($|m|=2-80$) Lindblad torque exerted on Saturn at different time points. It is
clear that during the first $300P_{S0}$ evolution, the total outer Lindblad
torque is greatly reduced to the same value as the inner one. Second, the gas
outside Saturn is pushed inward by the tidal torques of Jupiter. Parts of the
gas flow across Saturn's orbit and generate an additional positive corotation
torque on Saturn, which drives Saturn to move outward(see \textbf{Figure}
\ref{s005}). Third, the disk between Saturn and Jupiter is heavily weakened by
the gap created by Jupiter. As Jupiter approaches Saturn, more higher
order($|m|>2$) inner Lindblad torques of Jupiter and outer Lindblad torques of
Saturn are cleared. Thus, both parts of the disk, outside Jupiter and inside
Saturn, push the two planets toward each other further. Finally, when the disk
between them is swept out gently and a common gap forms finally(see
\textbf{Figure} \ref{d005}).

Since this convergent migration is slow, the $2:1$ MMR is then a robust
outcome. When Jupiter captures Saturn into MMR, Saturn is forced to migrate
inward to the vicinity of the central star, preserving a high eccentricity
$e_S\sim0.4-0.5$(\textbf{Figure} \ref{m05}). This result is consistent with the
observations, e.g. Gliese 876 and HD 128311. Faster convergent migration and
higher MMR may be achieved by reducing the initial separation between the two
planets further.

\begin{figure}
\epsscale{1} \plotone{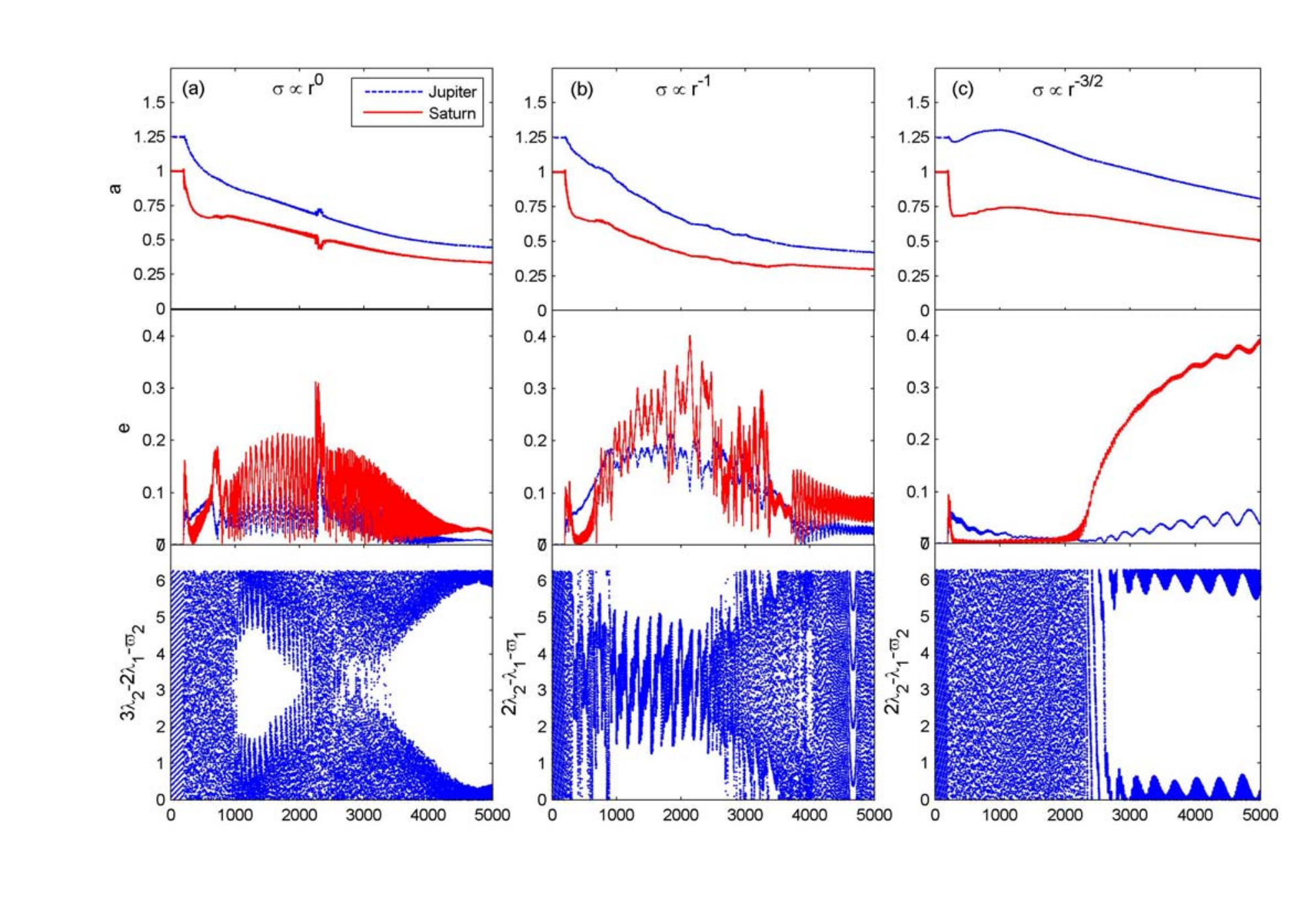} \caption{orbital
evolutions of Saturn and Jupiter embedded in a gas disk whose surface density
slope $\alpha$ is $0$(Panel a), $1/2$(Panel b) and $3/2$(Panel c). The initial
separation between the two planets is reduced to $d=0.25$. The convergent
migration becomes faster at small initial separation and the $3:2$ MMR is
reached when $\alpha=0$(Panel a). \label{m025}}
\end{figure}

At last, we try a small separation $d=0.25$ by setting Jupiter initially at
$a_{J0}=1.25$ and Saturn initially at $a_{S0}=1$. Since $d<3r_{mH}$, the mutual
interaction due to gravity between the giant planets becomes important at the
beginning of the evolution. Our results show that, although the steep surface
density prevents fast inward migration, it does not determine the direction of
the migration as in the previous cases. Now, the two planets migrate inward
even when $\alpha=3/2$. The results are similar to the cases that have $d=0.5$
and $\alpha<1/2$, but the inward migration of the planet pair becomes much
faster and more unstable. As the result of this fast convergent migration,
Saturn is trapped into the $3:2$ MMR with Jupiter when the disk is flat
$\alpha=0$. For the other surface density slopes $\alpha\geq1/2$, the $2:1$ MMR
is still the preferred outcome. The eccentricity of Saturn never gets higher
than $0.3$ in the $3:2$ MMR, see \textbf{Figure} \ref{m025}.

\subsection{Comparison between $q>1$ and $q<1$}
As we have shown in the previous section and in Paper I, both $q>1$ and $q<1$
configurations could result in convergent migration of the two planets and the
the planets being trapped in MMRs. The observations show that more than half of
the exoplanet pairs trapped in MMRs are of $q<1$ configuration. It should be
interesting to compare the results of these two configurations.

First, the major difference between the cases of $q>1$ and $q<1$ is the
direction of the common migration of the planet pair after they are trapped
into MMRs. When $q>1$, the common inward migration is halted and the planet
pair starts to migrate outward. When $q<1$, common inward migration is
preferred, see \textbf{Figures} \ref{m05} and \ref{m025}.

Since Jupiter is much more massive, Saturn in fact follows the migration of
Jupiter when they are locked in MMRs. The migration of Jupiter is dominated by
the torque balance between the gas disk inside and outside its orbit. Thus, the
surface density slope of the gas disk is an important issue(presented in paper
I). For a density slope that could result in convergent migration, e.g.
$\alpha=0$, the relative position of the two planets makes a difference: when
Saturn lies outside Jupiter ($q>1$), the total outer Lindblad torque exerted on
Jupiter is weakened and the inner Lindblad torque becomes relatively stronger.
Thus the inward migration of Jupiter will be slowed down, halted or even
reversed; when Saturn lies inside Jupiter ($q<1$), the situation is also
similar: the total inner Lindblad torque exerted on Jupiter is reduced when
Saturn sweeps out the gas inside Jupiter and the net inner Lindblad torque
exerted on Jupiter decreases. So, Jupiter will migrate inward and push Saturn
inward as well.

Second, the eccentricity of Saturn could be excited to a higher value in the
$q<1$ configuration. When Saturn lies outside Jupiter ($q>1$), its maximum
eccentricity could be excited to $e_S\sim0.2$ by the $2:1$ MMR and to
$e_S\sim0.15$ by the $3:2$ MMR. When $e_S>0.15$, the $3:2$ MMR breaks, and the
eccentricity is damped by gas disk quickly. Due to the damping effect of gas,
the eccentricity of Saturn never increases to $e_S>0.2$ even in $2:1$ MMR.
However, when Saturn lies inside Jupiter, its maximum eccentricity could be
excited to $e_S\sim0.4-0.5$ by the $2:1$ MMR and the system is still
stable(\textbf{Figure} \ref{m05}).

This difference could be understood by considering the different structures of
the heavily perturbed disk. When Saturn is located outside Jupiter($q>1$), the
gas disk would be separated into two parts after the two gaps overlap. As the
eccentricity keeps growing, Saturn gets closer and closer to the out edge of
the common gap or even cuts into the disk. If the resonance is strong enough,
e.g. the $2:1$ MMR, the effect of damping and excitation will allow the system
to achieve equilibrium. Otherwise the MMR would probably break at high
eccentricity, e.g. in the $3:2$ MMR(see Paper I). The situation seems to be the
same in the $q<1$ configuration since Saturn will meet the inner edge of the
common gap. However, since the two planets tend to migrate inward together
after the common gap is formed, Saturn usually forms an inner cavity at the
very center(\textbf{Figure} \ref{d005}). With both Saturn and Jupiter located
within the cavity, the damping effect of the gas is negligible. Thus the
eccentricity of Saturn could be estimated by mutual dynamical analysis.
Denoting $e_{S0}$, the eccentricity of Saturn before $p+1:p$ resonance, and
$a_{J0}$, $a_{J}$ the semimajor axes of Jupiter before and after resonance
respectively, we can obtain the eccentricity of Saturn $e_{S}$ which is excited
by the resonance through the following equation\citep{Mal95}:
\begin{equation}
e_S^2=e_{S0}^2+\frac{1}{p+1}ln(\frac{a_{J0}}{a_{J}}).
\end{equation}
For the $2:1$ MMR, when $e_{S0}=0.01$, $a_{J0}=1.05$ at $T\approx1000P_{S0}$
and $a_{J}=0.58$ at $T\approx5000P_{S0}$, both the above equation gives
$e_S=0.54$, and our result gives $e_S=0.5$, which are in good agreement with
each other(see \textbf{Figure} \ref{m05}).

The third difference is the frequency and stability of the planet pair in MMRs.
As shown here and in Paper I, the two planets undergo convergent migration more
easily in the $q>1$ configuration. Convergent migration and MMRs are robust
outcomes for all the surface density slopes when $q>1$. However, when $q<1$,
the two planets are under divergent migration in almost half of the
simulations.

\begin{figure}
\epsscale{1} \plotone{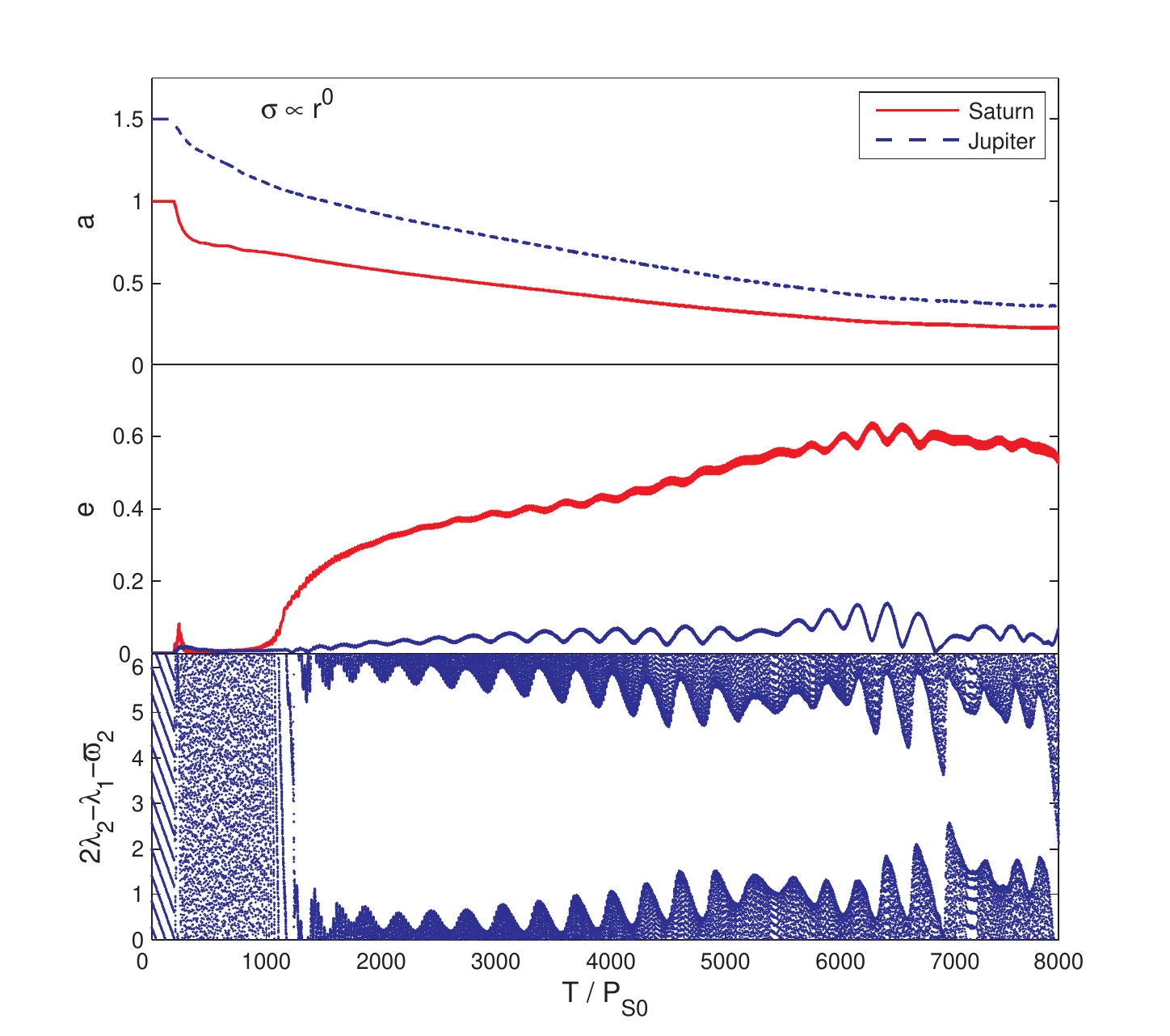} \caption{long time evolution
of $\alpha=0$ and $d=0.5$ case. This system is stable for at least $8000P_{S0}$
at high eccentricity of Saturn ($e_S\leq0.6$) when the $2:1$ MMR is preserving.
Saturn get close to the central star at $a_S<1AU$. \label{longtime}}
\end{figure}

It seems that, most of the exoplanet pairs trapped in the MMRs should be of the
$q>1$ configuration(where the inner planet is more massive than the outer one).
However, by considering the above-mentioned differences and the migration
process of the two planets, we find that the MMRs in $q<1$ are more stable than
that in $q>1$. In the case of $q>1$, Saturn and Jupiter usually reverse their
migration to outward when they are locked into MMRs. As Jupiter moves outward,
Saturn is pushed outward further and their orbits are in fact divergent. As the
separation between them increases, the mutual interaction due to gravity
becomes weaker and Saturn is easily scattered away by the corotation torque,
especially at a high eccentricity(unless the gas has already depleted). For
$q<1$, there is no such a problem since the two planets migrate inward and form
an inner cavity at the center. Our long time evolution also proves this
stability ($T\geq8000P_{S0}$, see \textbf{Figure} \ref{longtime}). This
explains why the relatively high frequency of exoplanet systems have the
configuration of $q<1$ contain MMRs; see \textbf{Table} 1. This relatively high
frequency may also be a result of observational bias where the massive
outermost planet is easier to detect by the radial velocity method.

\section{Discussions}

In our simulations, the onset of convergent migration is much earlier than the
emergence of the inner cavity. Therefore, in our results, the cavity should not
account for the suppression of the inward migration of Saturn. We would like to
indicate that there are two essentials that make convergent migration happen:
the steepened density slope on the inner disk(compressed by the pre-formed
giant planet) and the proper separation between Saturn and Jupiter.

The disk discussed here should be gas-full, otherwise it could not support the
long range type II migration of massive planets. This usually happens at the
early stage of the evolution of a protostellar disk. When a massive planet is
forming, say Jupiter, it will substantially change the structure of the gas
disk by digging a clear gap. The gas interior its orbit will be pushed inward
and accumulates on the inner disk. If the timescale of gas accretion onto the
central star(and other processes leading to loss of gas at the center) is
longer than the timescale of this compression, the local density slope on the
inner disk will steepen. The typical gas accretion rate of T Tauri star is
around $\sim10^{-7}M_\odot yr^{-1}$, while the gap opening process is only
$\sim100$ orbits if we drop a mature Jupiter in an unperturbed disk(in fact, we
have already employed a 'quiet start' method in our simulations to simulate the
growth of planets and avoid the unreal initial impact to the gas. The planets
all start with a mass of $0.1M_{\oplus}$ Earth masses, see section 2.2 and
paper I).

If we take into account the growth of the planet(core and gas accretion), the
gap opening process will be prolonged. However, according to the core accretion
model of giant planet formation, for the core mass stays below $15M_{\oplus}$
most of the time. Since the lowest mass required to open a gap in a typical
protostellar disk($H/r=0.04, \nu=10^{-5}$) is around $30-50M_{\oplus}$, the
planet would not change the disk significantly during this stage. As soon as
its core mass reaches the critical mass $M_{core}=15M_{\oplus}$ and equals to
the mass of its gas envelope, the gas accretion steps into the runaway
accretion phase\citep{Pol96} and the total mass of planet would increase to 1
Jupiter mass within several hundred orbits\citep{Ang08}. That means, once the
planet is massive enough to dig a gap on the disk, the gap will be opened and
enlarged very quickly. So the gas accumulation on the inner disk should be a
natural outcome when Jupiter is forming within the disk.

Besides the surface density factor, a proper separation between the two giant
planets is also required. If the separation is small, the gas disk between
Saturn and Jupiter would be weakened substantially when the two planets are
digging gaps on the disk. For Jupiter(Saturn), the torques that come from the
gas inside(outside) its orbit are reduced and it would be pushed
inward(outward) by the disk outside(inside) its orbit. And, as the two planets
get closer, the torque unbalance of each planet becomes more serious and the
two planets would get closer. Thus, the total effect is that the gas disk tends
to keep pushing the two giant planets toward each other when the gas between
their orbits becomes more and more tenuous. When the two planets get close
enough(the gaps usually had already overlapped), the mutual gravitational
interaction between the two planets would prevent them from further
approaching, e.g. the MMRs.

Our results show that the maximum separation leading to convergent migration is
approximately $d\leq5r_{mH}$(when $\alpha=0$). According to the core accretion
model, the embryos of planet will accrete all the solid mass within their
vicinities and achieve their isolation masses. This isolation separation
between embryos is also $d_{iso}\sim5r_{mH}$. If we take into account of the
migration of light planets(embryos) in the gas disk, the planets may get closer
before they become massive enough to open gaps on the disk. Furthermore, the
massive gas giants usually form in an area a little further outside the snow
line, where the surface density of disk increases significantly due to the
icing of water, and the collision timescale of solid grains is not long enough
to prevent the effective growth of the planet core. So, it is reasonable to
expect that some giant planet cores emerge with proper separations. This means
that the initial conditions leading to convergent migration should not be rare,
despite $q>1$ or $q<1$($q$ is the mass ratio of the inner planet and the outer
planet).

To focus on the effects of the surface density slope of the disk and the
initial separation between the two planets, we assume the two planets form
simultaneously and the gas accretions onto planets(and the growth of planet)
are not taken into account in this series of works(in both paper I and II).
However, this does not mean that the accretion process is negligible. In fact,
the gas accretion should become more vigorous on Saturn, whose gap formation
process is well prolonged by Jupiter. Thus, we add some additional discussions
about the accretion processes of the planets here. Many factors need to be
considered when the mass growing of the planet is included, e.g. the formation
sequence of the two planets, the time required to build a critical mass core,
and the proper descriptions of accretion rate and range. We find that it
becomes too complicated to concentrate on the dynamic evolution of the planet
pair if all the factors are taken into account. To avoid the complexity, one
could take a reasonable assumption that Jupiter forms first while Saturn is
still a planet core undergoing gas accretion. Although, in a multiple planet
system, the accretion process of planet makes the orbit evolution more
complicated by leading to mass of the planets growing and various migration
rates, we believe that the it may not change our main result---the convergent
migration---qualitatively.

First of all, the accretion onto Jupiter could be neglected since the clear gap
prevents the effective gas accretion process. Then we only need to consider the
accretion onto Saturn. One could further assume the core of Saturn is already
around $15M_\oplus$ and it undergoes fast type I migration initially. Since the
direction of type I migration is always(or in most cases) inward, the
convergent migration should still be a robust outcome when the planet
core(Saturn) lies outside Jupiter($q>1$). Comparing to the cases we studied in
this paper, the growing Saturn may get closer to Jupiter and be trapped in MMRs
with higher $p$($p+1:p$ is the orbit ratio of the two planets). As the mass of
Saturn core keeps increasing, the high-$p$ MMRs would become unstable and lead
to breaks or re-captures of MMRs(the resonances may overlap easily for massive
planets at high $p$ and leads to instability, see paper I).

When the planet core lies inside Jupiter($q<1$), the two planets would undergo
divergent migration at the beginning. However, the accretion process is fully
runaway when the core mass is above $15M_\oplus$. It will take only $~200-400$
orbits to achieve $1M_J$ for an accreting planet embedded in a disk as dense as
the one we adopt here and its orbit decay is less than $~20\%$ during this
process\citep{Ang08}. When the mass of the inner planet becomes massive enough
to open a clear gap, its migration rate will decrease significantly.
Furthermore, as it keeping accreting gas, the inner planet should become more
massive(than Saturn) and dig a wider gap, which weakens the disk between the
two planets further. Thus, the gas outside the planet pair(inside inner planet
and outside outer planet) tends to push the two planets toward each other(the
inner planet only gets angular momentum from inner disk and move outward, the
outer planet only loses angular momentum to outer disk and move inward) and
results in a more compacter orbital configuration. So, it may also lead to the
convergent migration in the $q<1$ configuration, when we consider the gas
accretion onto the Saturn core.

A more self-consistent simulation should include two accreting and interacting
planet cores as well as the thermal evolution of the gas around planets, which
would lead to much more complicated orbital evolution and thus stepping into a
new aspect of orbital evolution investigation. The relational results are under
preparation now.

We also note that the characteristics of resonant systems with $q<1$ are the
relatively large eccentricity and the short orbit period of the inner giant
planet, e.g. GJ 876, HD 128311, HD 45364 and HD 60532. These characteristic
orbits are easier to detect by radial velocity methods. Thus, this
observational bias also explains the relatively high frequency of the resonant
system in the $q<1$ configuration. The orbital eccentricities of the planets in
these real systems are all less than $0.3$, which indicates that there may
exist other mechanisms that restrain the planets' eccentricities when the gas
damping is absent. One possible mechanism which need to be addressed further is
the interaction with planetesimal disk after the depletion of gas.

The tidal dissipation that arise from the star will drive the resonant planet
pair out of resonance, when they are close to the center(D.N.C. Lin et al.
2010, in preparation). The eccentricity of the inner planet could be damped by
this tidal dissipation as well. So the 'hot Jupiter/Saturn' would probably form
in the $q<1$ system as follows: at first, the fast inward migration of the
inner planet is suppressed by the outer giant planet. Then the two planets are
locked into MMRs and migrate inward together. They dig an inner cavity at the
center of gas disk, and the eccentricity of inner planet is excited to a high
value by the resonance. As the inner planet gets closer to the star, the tidal
dissipation gently drives it out of the MMR with the outer one. After the
depletion of gas, the inward migration of outer planet stops, while the inner
one falls toward the center due to tidal interaction with the star. Finally,
the two planets undergo divergent migration again and the inner one approaches
the central star more closely with a moderate eccentricity which had been
damped by the tidal dissipation of the star. To ensure the validity of this
process, we need to carry out further N-body integration associated with tidal
damping, by adopting the results of this work as the initial conditions.

\section{Conclusions}
Following the paper I, we continue the investigation of the orbital evolution
of Saturn and Jupiter embedded in a protostellar disk by running a series of 2D
high-resolution hydrodynamic simulations. The main aim of this work is to find
out whether the convergent migration also happens in a system with $q<1$(where
the more massive giant planet is initially located outside the lighter one). To
do so, we switch the initial positions of Saturn and Jupiter to achieve $q<1$
and focus on the effects of various surface density profiles of the gas disk
and different initial separations between the two planets. From our results and
analysis, we summarize our conclusions as follows:

(1). The type II migration of Saturn could be suppressed by Jupiter when $q<1$.
As Jupiter digs a deep gap, the gas disk is cut into two parts
---an inner disk and an outer disk. Saturn also digs a gap on the inner disk and
follows the viscous movement of gas. Being shepherded by the tidal torque of
Jupiter, the expansion of inner disk is suppressed. When the surface density
slope is steep $\alpha>1/2$, Jupiter migrates outward and the inner disk
expands. The width of gap increases as Jupiter moves outward($a_J$ increases)
and the expansion of inner disk is limited, thus the outward migration of
Saturn is limited as well. When the surface density is nearly flat
$\alpha\leq1/2$, Jupiter tends to migrate inward. Being compressed by Jupiter,
the inner disk becomes denser and the local surface density slope $\alpha$
increases. Thus the inner disk tends to spread outward and fight against the
tidal compression of Jupiter. When the expansion and compression effects
achieve equilibrium, the inward migration of Saturn is slowed down or even
halted.

In a system of $q<1$, this mechanism provides a way to halt the inward
migration of the inner giant planet and does \emph{not} require the overlapping
of gaps. In fact, this suppression happens as early as the outer massive planet
starts to compress the inner disk and makes the convergent migration possible.
We also note that, the inner planet should also be massive enough to open a gap
on the disk, otherwise it won't follow the viscous evolution of the inner disk
and this mechanism won't be valid. The lowest mass required should be
$30-50M_e$, depending on the scale height and viscosity of the disk.

(2). Convergent migration could also happen in $q<1$ configuration under some
circumstances. The two main factors that account for the occurrence of the
convergent migration: are the nearly flat surface density profile and the
relatively small separation between the two planets. On the one hand, as we
have concluded above, when $\alpha\leq1/2$, the inward migration of Saturn
would be suppressed by the steepened local surface density slope on the inner
disk. On the other hand, as the initial separation between the two planets is
reduced, the total inner(outer) Lindblad torque exerted on Jupiter(Saturn) is
weakened by the gap created by Saturn(Jupiter). Thus, the outer torque
overwhelms the inner one and pushes Jupiter inward faster than the regular type
II migration. In the meantime, the inward migration of Saturn is suppressed and
it migrates more slowly than Jupiter. As a result, the net orbital movement of
the two planets is convergent.

The $2:1$ MMR is a usual outcome when Jupiter and Saturn approach to each other
adiabatically. If the initial separation becomes smaller, e.g.
$d\leq0.25\sim3r_{mH}$, the inward migration of Jupiter is much faster and the
two planets may achieve the $3:2$ MMR. However, because of the mutual
scattering due to gravity, the migrations of both planets are unstable when the
initial separation is too small.

(3). In the case of $q<1$, after Saturn and Jupiter have been locked into MMRs,
they will migrate inward together instead of migrating outward, and the
eccentricity of Saturn could be excited much higher than that of the $q>1$
configuration. After the stage of convergent migration, the gaps of the two
planets overlap. As the disk between the two planets is cleared, Saturn is
pushed outward by the inner disk and Jupiter is pushed inward by the outer
disk. When the two approaching planets are locked into MMR, the mutual
interaction due to gravity prevents them from getting close too quickly, and
the separation between them becomes steady. Since Jupiter is much more massive,
Saturn is then pushed inward by Jupiter. As they move toward the center, the
inner disk is swept out and forms an inner cavity.

This cavity plays a great role. First, without the damping effect of gas, the
eccentricity of Saturn could be excited to $e_{S}\sim0.5$(in the $2:1$ MMR),
which is much higher than that of $q>1$ configuration and is consistent with
the result of the analysis of two resonant planets. Second, without the
scattering triggered by the massive gas within the coorbital zone of planet,
Saturn is able to maintain the high eccentricity and be pushed to the vicinity
of the central star. Our long-time evolution shows that the inward migration of
Saturn and Jupiter is stable at the high eccentricity for at least
$T\geq8000P_{S0}$. Thus, this cavity in fact ensures the stability of a highly
eccentric system, e.g. 'Hot Jupiter/Saturn'.

\section{Acknowledgement}
We thank D.N.C. Lin, A.Crida, W. Kley for their constructive conversations.
Zhou is very grateful for the hospitality of Issac Newton institute during the
Program `Dynamics of Discs and Planets'. This work is supported by NSFC (Nos.
10925313,10833001,10778603), National Basic Research Program of
China(2007CB814800) and Research Fund for the Doctoral Program of Higher
Education of China (20090091120025).

{}
\clearpage

\begin{deluxetable}{lccccccr}
\tablecaption{An incomplete list of giant exoplanets probably locked in low
order MMRs.\label{table 1}} \tablehead{\colhead{System} & \colhead{ No.} &
\colhead{$P$ [day]} & \colhead{$M\sin i$ [$M_J$]} & \colhead{$a$ [$AU$]} &
\colhead{$e$} & \colhead{$q=M_i/M_o$} &\colhead{Refs.}}\startdata
Gliese 876  & c & 30.34 & 0.619 & 0.13 & 0.22 & $\sim 0.31$ & 1\\
(2:1)       & b & 60.94 & 1.935 & 0.21 & 0.02 &   \\
\tableline
HD 73526& b & 188 & 2.90 & 0.66 & 0.19 & $\sim 1.16$ & 2\\
(2:1)   & c & 377 & 2.50 & 1.05 & 0.14 &  \\
\tableline
HD 82943& c & 219 & 2.01 & 0.75 & 0.36 & $\sim 1.14$ & 3\\
(2:1)   & b & 441 & 1.75 & 1.19 & 0.22 &   \\
\tableline
HD 128311& b & 448 & 2.18 & 1.10 & 0.25 & $\sim 0.67$ & 4\\
(2:1)    & c & 919 & 3.21 & 1.76 & 0.17 &   \\
\tableline
HD 160691& d & 310 & 0.52 & 0.92 & 0.07 & $\sim 0.31$ & 5\\
(2:1)    & b & 643 & 1.68 & 1.50  & 0.13 &  \\
\tableline
47 Uma   & b & 1083 & 2.6  & 2.11  & 0.05 & $\sim 5.65$ & 6\\
(2:1)    & c & 2190 & 0.46 & 3.39  & 0.22 &   \\
\tableline
HD 45364& b & 227 & 0.19 & 0.68 & 0.17 & $\sim 0.29$ & 7\\
(3:2)   & c & 343 & 0.66 & 0.90 & 0.10 &   \\
\tableline
HD 60532& b & 201 & 3.15 & 0.77 & 0.28 & $\sim 0.42$ & 8\\
(3:1)   & c & 605 & 7.46 & 1.58 & 0.04 &  \\
\tableline
55 Cnc  & b & 14.6 & 0.82 & 0.11 & 0.01 & $\sim 4.94$ & 9\\
(3:1)   & c & 44.3 & 0.17 & 0.24 & 0.09 &   \\
\enddata
\tablecomments{{\bf References.} (1) Marcy et al. 2001;(2) Tinney et al. 2006;
(3) Lee et al. 2006; (4) Vogt et al. 2005; (5) Gozdziewski et al. 2007;
(6)Guillem et al. 2010; (7)Correia et al. 2009; (8) Laskar \& Correia 2009;
(9)Fischer et al. 2003. }
%% Text for table notes should follow after the \enddata but before
%% the \end{deluxetable}. Make sure there is at least one \tablenotemark
%% in the table for each \tablenotetext.
%%\tablecomments{Table 1. }
%%\tablenotetext{a}{Sample footnote for table~\ref{tbl-1} that was
%%generated with the deluxetable environment}
%%\tablenotetext{b}{Another sample footnote for table~\ref{tbl-1}}
\end{deluxetable}
\begin{deluxetable}{cccccc}
\tablecaption{A summary of our simulations.\label{table 2}} \tablehead{
\colhead{Case} & \colhead{Configuration} & \colhead{Separation $d$} &
\colhead{$\sigma$} & \colhead{Relative migration} & \colhead{Resonance} }
\startdata
1  & S-J & $d=1$ & $\sigma \sim r^0$ & Divergent & - \\
\tableline
2  & S-J & $d=1$ & $\sigma \sim r^{-1/2}$ & Equilibrium & -  \\
\tableline
3  & S-J & $d=1$ & $\sigma \sim r^{-1}$ & Divergent & -  \\
\tableline
4  & S-J & $d=1$ & $\sigma \sim r^{-3/2}$ & Divergent & -  \\
\tableline
5  & S-J & $d=0.5$ & $\sigma \sim r^0$ & Convergent & $2:1$ \\
\tableline
6  & S-J & $d=0.5$ & $\sigma \sim r^{-1/2}$ & Convergent & $2:1$  \\
\tableline
7  & S-J & $d=0.5$ & $\sigma \sim r^{-1}$ & Divergent & -  \\
\tableline
8  & S-J & $d=0.5$ & $\sigma \sim r^{-3/2}$ & Divergent & -  \\
\tableline
9  & S-J & $d=0.25$ & $\sigma \sim r^0$ & Convergent & $3:2$ \\
\tableline
10 & S-J & $d=0.25$ & $\sigma \sim r^{-1/2}$ & Convergent & $2:1$  \\
\tableline
11 & S-J & $d=0.25$ & $\sigma \sim r^{-1}$ & Convergent & $2:1$  \\
\tableline
12 & S-J & $d=0.25$ & $\sigma \sim r^{-3/2}$ & Convergent & $2:1$  \\
\enddata
\end{deluxetable}

\end{document}